\documentclass[letterpaper, 10 pt, conference]{ieeeconf}  % Comment this line out

\usepackage[hidelinks]{hyperref}
\hypersetup{
    colorlinks=false,
    linkcolor=black,
    filecolor=magenta,      
    urlcolor=blue,
}
\usepackage{amsmath}
\usepackage{graphicx}
\usepackage{amssymb}
\usepackage{colortbl}
\usepackage{arydshln}
\usepackage{stfloats}
\usepackage[vlined,boxed,ruled,linesnumbered]{algorithm2e}
\SetKwProg{Fn}{Function}{}{}
\usepackage{algpseudocode}
\usepackage{multirow}
\usepackage{extarrows}
\usepackage{mathrsfs}
\usepackage{upgreek}
\usepackage{xcolor}

\newtheorem{remark}{Remark}

\newtheorem{problem}{Problem}

\newtheorem{assumption}{Assumption}

\newtheorem{definition}{Definition}

\newcommand{\weiming}[1]{{\textcolor{black}{#1}}}

\definecolor{mygray}{gray}{0.8}

\IEEEoverridecommandlockouts                              % This command is only
\title{\LARGE \bf \weiming{A Data-Driven Modeling Framework of Time-Dependent Switched Dynamical Systems via Extreme Learning Machine} 
}

\author{Weiming Xiang,~\IEEEmembership{Senior Member, IEEE}% <-this % stops a space
\thanks{}% <-this % stops a space
\thanks{Weiming Xiang is with the School of Computer and Cyber Sciences, Augusta University, Augusta GA 30912 USA. Email:
        {\tt\small wxiang@augusta.edu}}%%
}

\begin{document}

\maketitle
\thispagestyle{empty}
\pagestyle{empty}

%%%%%%%%%%%%%%%%%%%%%%%%%%%%%%%%%%%%%%%%%%%%%%%%%%%%%%%%%%%%%%%%%%%%%%%%%%%%%%%%
\begin{abstract}
\boldmath
In this work, \weiming{a data-driven modeling framework of switched dynamical systems under time-dependent switching is proposed.} The learning technique utilized to model system dynamics is Extreme Learning Machine (ELM). First, \weiming{a method is developed for the detection of the  switching occurrence events in the training data extracted from system traces.} The training data thus can be segmented by the detected switching instants. Then, ELM is used to learn the system dynamics of subsystems. The learning process includes segmented trace data merging and subsystem dynamics modeling. Due to the specific learning structure of ELM, the modeling process is formulated as an iterative Least-Squares (LS) optimization problem. Finally, the switching sequence can be reconstructed based on the switching detection and segmented trace merging results. An example of the data-driven modeling DC-DC converter is presented to show the effectiveness of the developed approach.
\end{abstract}

%%%%%%%%%%%%%%%%%%%%%%%%%%%%%%%%%%%%%%%%%%%%%%%%%%%%%%%%%%%%%%%%%%%%%%%%%%%%%%%%
\section{Introduction}
A switched system is composed of a family of continuous or discrete-time subsystems along with a switching rule orchestrating the activation of them. Switched systems are frequently used to model and characterize the behaviors of practical systems that are inherently multi-modal such as power electrical systems, medical robotic systems, autonomous vehicles, and a variety of Cyber-Physical Systems (CPS). Relevant analysis, control, and verification methods and tools for hybrid systems were developed based on explicit models.  For instance, multiple Lyapunov function approaches were developed for stability analysis for switched systems and extended to a variety of related problems \cite{lin2009stability,liberzon1999basic,xiang2018nonconservative,xiang2016necessary}.  After several decades of development of switched and hybrid system theory and increasingly implementations to various applications, switched system theory shows its powerful capability in dealing with complex dynamical systems, and those achievements are made on the basis of explicit mathematical models. However, accurate models of many practical systems such as complex CPS are often not easy (or even impossible) to obtain due to incomplete knowledge of certain processes. 

Data-driven methods have been shown to outperform physics-based models in many disciplines, especially for those physical processes that are not fully understood by researchers, but for which data of adequate quality and quantity is available. When applying data-driven modeling methods for switched systems, there are three unique challenging problems that need to be taken into account:
\begin{itemize}
    \item \emph{How does one detect the switching occurrence and segment collected system traces/data?}
    \item \emph{How does one classify and model multiple subsystems?}
    \item \emph{How does one reconstruct switching signals?}
\end{itemize}

The above three core problems are the main focus of data-driven modeling of switched systems. In \cite{ohlsson2013identification}, a general convex framework for the identification of switched linear systems is proposed in the framework of the Least-Squares (LS) problem with sum-of-norms regularization. A model identification framework is proposed for hybrid systems consisting of linear subsystems in \cite{medhat2015framework}, in which linear regression is used to estimate the dynamics. In \cite{zhang2016switched}, the switched system identification problem is formulated as a constrained multi-objective optimization problem. A recursive procedure is developed in \cite{bako2011recursive} for the identification of switched linear models from input-output data. For more details of the results on the modeling and identification for switched systems and hybrid systems, the reader is referred to survey papers such as \cite{garulli2012survey,paoletti2007identification}, and the references cited therein. \weiming{This paper aims to propose a data-driven modeling framework of time-dependent switched systems by using a learning technique called Extreme Learning Machine (ELM) which is with a learning structure appropriate for modeling switched systems.} First, a switching detection method is developed to detect the occurrence of switchings and segment the collected system traces. Then, taking advantage of the ELM learning structure in the form of an LS optimization problem, an algorithm is developed to merge segmented traces and then model subsystem dynamics. Finally, the switching sequence can be reconstructed based on the switching detection and segmented trace merging results.  

The rest of this paper is organized as follows. The problem formulation and preliminaries are given in Section II. \weiming{A data-driven modeling framework of switched systems using ELM are presented in Section III.} In Section IV, an evaluation is performed on the modeling of DC-DC converters. \weiming{Conclusion and further remarks are given in Section V.}

\section{Preliminaries and Problem Formulation}
%\subsection{Switched System Model}
In this paper, the dynamical systems considered for modeling are assumed to be a class of switched nonlinear systems in the following  form
\begin{equation}\label{NARMA}
\mathbf{x}(k+1)=f_{\theta(k)}\left(\mathbf{x}(k),\mathbf{u}(k)\right)
\end{equation}
where $\mathbf{x}(k)\in \mathbb{R}^{n_x}$ is the system state, and $\mathbf{u}(k)\in \mathbb{R}^{n_u}$ is the system input, respectively.  Defining index set $\mathcal{I} \triangleq \{1, 2,\ldots, N\}$ where $N$ is the number of subsystems, $\theta(k): \mathbb{N} \to \mathcal{I}$ is a function of time $k$ denoting the switching signal. $f_i$, $i \in \mathcal{I}$ is the unknown nonlinear discrete-time process for the $i$th subsystem, characterizing system dynamical behaviors. However, accurate models of many practical systems are often not easy (or even impossible) to obtain due to incomplete knowledge of certain processes including switched nonlinear systems with multiple subsystems along with a switching law. \weiming{In this paper, we are going to develop a data-driven modeling framework to construct individual subsystem dynamics $f_i$, $i\in \mathcal{I}$ and switching law $\theta(k)$ using the data collected from system trajectories and inputs, i.e., the input-state traces $\{\mathbf{x}(k),\mathbf{u}(k)\}$, $k=0,1,\ldots,K$.}

\begin{definition}
A input-state trace $\phi_K$ is a sequence of pairs of $\psi(k)=\{\mathbf{x}(k),\mathbf{u}(k)\}$, $k=0,1,\ldots,K$ generated by system (\ref{NARMA}) starting from initial values $\psi(0)=\{\mathbf{x}(0),\mathbf{u}(0)\}$.
\end{definition}

In this paper, it is assumed that there exist sufficient measurable input-state traces available for the data-driven modeling for switched nonlinear systems in the form of  (\ref{NARMA}). 

\begin{assumption} \label{assumption_1}
It is assumed that state $\mathbf{x}(k)$ and input $\mathbf{u}(k)$ are all measurable for data-driven modeling of switched nonlinear system (\ref{NARMA}). 
\end{assumption}

\begin{remark}
\weiming{
In practice, the collected data inevitably contains noise, e.g., measured state $\hat{\mathbf{x}}(k) = \mathbf{x}(k) + \delta(k)$, where $\delta(k)$ is the noise vector at instant $k$.  There are various noise reduction algorithms available to reduce or remove noise from a signal. These denosing methods heavily rely on the types of signals and noise, and multiple methods are successful within the scope of their target signal and noise type. In the rest of the paper, we assume that  collected traces are well pre-denoised using some 
noise reduction method that is not in the scope of this work, and we still use $\mathbf{x}(k)$ and  $\mathbf{u}(k)$ to denote denoised measured state and input, if there is no ambiguity in the context. }
\end{remark}

%There exist various data-driven learning techniques. To meet the complexity of switched systems with the unique feature of multiple modes and switching laws orchestrating between them. In this paper, we will utilize the Extreme Learning Machine (ELM) learning technique, a class of neural network with a simple single-hidden layer structure but offering sufficient learning capability for the data-driven modeling of switched nonlinear dynamical systems in the form of (\ref{NARMA}).  

%\subsection{Extreme Learning Machine}
Extreme Learning Machine (ELM) is dedicated to train Single-hidden Layer Feedforward Networks (SLFN). An SLFN with $L$ hidden nodes can be represented in the form of
\begin{align}\label{SLFN}
\mathbf{y}=f(\mathbf{x}) = \sum\nolimits_{i=1}^{L}\boldsymbol\upbeta_ih_i(\mathbf{x}) = \mathbf{h}(\mathbf{x})\boldsymbol {\upbeta}
\end{align}
where $\mathbf{x} \in \mathbb{R}^{n}$ and $\mathbf{y} \in \mathbb{R}^{m}$ are the input and output vectors, respectively. The output nodes are chosen linear and $\boldsymbol{\upbeta} \triangleq [\boldsymbol\upbeta_1,\ldots,\boldsymbol\upbeta_{L}]^{\top}$ is the output weight vector between the hidden layer of $L$ nodes to the $m \ge 1$ output nodes. $\mathbf{h}(\mathbf{x})\triangleq [h_1(\mathbf{x}),\ldots,h_L(\mathbf{x})]$ is the ELM nonlinear feature mapping. In particular, $h_i(\mathbf{x})$ can be defined as follows:
\begin{align}\label{activation}
h_i(\mathbf{x}) = g(\mathbf{a}_i\cdot\mathbf{x} + b_i),~\mathbf{a}_i \in \mathbb{R}^d,~b_i \in \mathbb{R}
\end{align}
where $g(\cdot)$ is the activation function for hidden layer neurons including the sigmoidal functions as well as the radial basis, sine, exponential,
and many other nonregular functions.  $\mathbf{a}_i$  and $b_i$ are the  the input weight vector and the bias weight of the $i$th neuron in hidden layer. 

Given $N$ arbitrary distinct input-output samples $\{\mathbf{x}_i,\mathbf{t}_i\}$ with $\mathbf{x}_i \in \mathbb{R}^{n}$ and $\mathbf{t}_i \in \mathbb{R}^{m}$, ELM aims to find the following minimum norm Least-Squares (LS) solution of SLFN, which is expressed by
\begin{align}\label{elm_opt}
    \min_{\boldsymbol \upbeta \in\mathbb{R}^{L \times m}}\left\|\mathbf{H}\boldsymbol \upbeta - \mathbf{T}\right\|
\end{align}
where $\mathbf{H}$ is the randomized hidden layer output matrix as
\begin{align} \label{H}
   \mathbf{H} = 
   \begin{bmatrix} 
   \mathbf{h}(\mathbf{x}_1) \\
   \vdots
   \\
   \mathbf{h}(\mathbf{x}_N) 
   \end{bmatrix} =
   \begin{bmatrix} 
h_1(\mathbf{x}_1)  & \cdots & h_L(\mathbf{x}_N) \\ 
\vdots & \ddots & \vdots \\ 
h_1(\mathbf{x}_1)  & \cdots & h_L(\mathbf{x}_N) 
\end{bmatrix}
\end{align}
and $\mathbf{T}$ is the training data target matrix
\begin{align}  \label{T}
    \mathbf{T} = 
    \begin{bmatrix}
    \mathbf{t}_1^{\top}
    \\
    \vdots
    \\
    \mathbf{t}_N^{\top}
    \end{bmatrix}
    =
    \begin{bmatrix}
    t_{11} & \cdots & t_{1m} \\
    \vdots & \ddots & \vdots \\ 
    t_{N1} & \cdots & t_{Nm} 
    \end{bmatrix}.
\end{align}
\begin{remark}
\weiming{
Compared with other conventional iterative learning frameworks, this ELM algorithm can provide a good generalization performance as well as an extremely fast learning speed. Even with one single randomized hidden layer, ELM is still a learner with universal approximation and  classification capabilities as proved in \cite{huang2006universal}. This universal approximation feature make ELM be able to capture the nonlinear dynamics in switched systems.
}
\end{remark}

\begin{remark}\weiming{
According to \cite{huang2006extreme,banerjee1973generalized}, the the smallest norm LS solution $\boldsymbol \upbeta^{*}$ for optimization problem (\ref{elm_opt}) is 
\begin{align}\label{elm_solution}
{\boldsymbol \upbeta^{*}} = \mathbf{H}^\dagger\mathbf{T}
\end{align}
where $\mathbf{H}^\dagger$ is the Moore-Penrose generalized inverse of matrix $\mathbf{H}$. Remarkably, the input weight and bias for hidden layer do not need to be tuned  once random values have been assigned at the beginning of learning, and the learning process only requires the computation on $\boldsymbol\upbeta^{*}$, which is extremely efficiently described by (\ref{elm_solution}). }
\end{remark}

%Thus, ELM offers sufficient learning ability for dynamical systems and, more importantly, the optimization-based learning process provide a unique learning framework for switched dynamical systems with multiple modes.

%The basic ELM algorithm for SLFN is proposed in \cite{} which is summarized in Algorithm \ref{alg1}.
%\begin{algorithm}[ht!]
%\SetAlgoLined
%\SetKwInOut{Input}{Input}
%\SetKwInOut{Output}{Output}
%\SetKw{Return}{return}
%\Input{Input-output samples $\{\mathbf{x}_i,\mathbf{t}_i\}$,  activation function $g(\cdot)$, hidden node number $L$}
%\Output{Weights $\mathbf{a}_i$ and bias $b_i$ for hidden layer, output weight $\beta_i$ for output layer}
%\Fn{$\mathrm{elm}(\{\mathbf{x}_i,\mathbf{t}_i\},~g(\cdot),~L)$}{
%Randomly assign weight $\mathbf{a}_i$ and bias $b_i$, $i = 1,\ldots,L$;

%Calculate the hidden layer output matrix $\mathbf{H}$ by (\ref{H});

%Calculate the output weight $\boldsymbol \upbeta$ by solving (\ref{elm_opt}), e.g., by (\ref{elm_solution}).
%}
% \caption{Basic ELM algorithm} \label{alg1}
%\end{algorithm}

%\subsection{Problem Formulation}

\weiming{Under Assumption \ref{assumption_1} and assuming that all the collected traces are with a same initial time and a same length $K$, we assume that there exist $N$ input-state traces defined by}
\begin{equation}
    \phi_K^{(n)}= \{\psi(0),\psi(1),\ldots,\psi(K)\},~n=1,2,\ldots,N
\end{equation}
which are collected from the system. The main objective is to extract switched system model (\ref{NARMA}) including subsystem dynamics $f_i$, $i \in \mathcal{I}$ and switching law $\theta(k)$. In the framework of ELM, the subsystem dynamics $f_i$, $i \in \mathcal{I}$ are supposed to be modeled in the form of (\ref{SLFN}). To achieve the data-driven modeling goal of switched nonlinear system (\ref{NARMA}), three core sub-problems need to be solved, i.e., switching instant detection, subsystem modeling, and switching law reconstruction. Thus, the data-driven modeling problem of switched systems using ELM is summarized as below:

\begin{problem} \label{problem_1}
Given $N$ input-state traces $\phi_K^{(n)}$, $n=1,2,\ldots,N$ collected from the system, how does one extract the switched system model (\ref{NARMA}), including the following three sub-problems:
\begin{enumerate}
    \item How does one determine the set of switching instants $\mathcal{S} =\{k_1,k_2,\ldots,k_\ell\}$, $0 < k_1 \le\ k_\ell < K$?
    \item How does one model subsystem dynamics $f_i$, $i \in \mathcal{I}$ by ELM? 
    \item How does one reconstruct time-dependent switching law $\theta(k)$?
\end{enumerate}
\end{problem}

Problem \ref{problem_1} with the three core problems is the main concern to be addressed in the rest of this paper.

\section{\weiming{Data-Driven Modeling Framework}}
In this section, a  data-driven modeling framework of switched systems will be presented, including switching detection, subsystem modeling, and switching reconstruction.

\subsection{Switching Detection}

\weiming
{
The occurrence of switching is usually reflected by the abrupt change of state trajectories. First, let's consider a continuous-time switched nonlinear system
\begin{equation}
    \dot{\mathbf{x}}(t) = f_{\theta(t)}(\mathbf{x}(t),\mathbf{u}(t))
\end{equation}
where  $f_i$, $i \in \mathcal{I}$ are $\mathcal{C}^{\infty}$ functions. The set of switching instants is denoted by $\mathcal{S} = \{t_1,t_2,\ldots\}$. To detect switching instants in $\mathcal{S}$, the key is to identify the events that indicate the occurrence of switchings. 
%\noindent  \textbf{Switching Occurrence Event:}
\begin{definition}
A switching occurrence event is an outcome that can be only resulted from the occurrence of switching.
\end{definition}
}

\weiming
{
Under the trivial assumption that input signals $\mathbf{u}(t)$ are assumed to be  $\mathcal{C}^{0}$ functions and due to the continuity of $f_i$, the following condition explicitly implies an switching occurrence event at time instant $t$:
\begin{align}
    \lim\nolimits_{t \to t^{-}}f_{\theta(t)}(\mathbf{x}(t),\mathbf{u}(t)) \ne     \lim\nolimits_{t \to t^{+}}f_{\theta(t)}(\mathbf{x}(t),\mathbf{u}(t))  \label{eq:switching_f}
\end{align}
which can be reflected by the measured state $\mathbf{x}(t)$ such that 
\begin{align}
    \left\|\mathbf{x}'_{+}(t) - \mathbf{x}'_{-}(t)\right\| > 0 \label{eq:switching_x}
\end{align}
where $\mathbf{x}'_{-}(t)\ = \limsup\nolimits_{h \to 0^{+}} \frac{\mathbf{x}(t) - \mathbf{x}(t-h)}{h}$ and $\mathbf{x}'_{+}(t) = \limsup\nolimits_{h \to 0^{+}} \frac{\mathbf{x}(t+h) - \mathbf{x}(t)}{h}$. Thus, the switching set determined by switching occurrence event (\ref{eq:switching_x}) is $\mathcal{S}' = \{t \in \mathbb{R}_+ \mid \left\|\mathbf{x}'_{+}(t) - \mathbf{x}'_{-}(t)\right\| > 0 \}$, and we have $\mathcal{S}' \subseteq \mathcal{S}$.
}

\weiming
{
It worth mentioning that (\ref{eq:switching_f}) and (\ref{eq:switching_x}) may not be able to characterize all switching occurrence events, e.g., if we consider a switched system with two modes $f_i(\mathbf{x}(t))$, $i=1,2$, and the event (\ref{eq:switching_x}) does not happen if the system switches at a particular state in set $\mathcal{X} = \{\mathbf{x} \mid f_1(\mathbf{x}) = f_2(\mathbf{x}) \}$. In this case, we can further exploit the continuity of the derivative of $f_i$ in addition to $f_i$. Given a trivial assumption $\mathbf{u}(t) \in \mathcal{C}^{1}$ and due to $f_i \in \mathcal{C}^{\infty}$, the following condition implies an switching occurrence event at time instant $t$:
\begin{align}
    \lim\nolimits_{t \to t^{-}}\frac{\partial f_{\theta(t)}}{\partial \mathbf{x}}&f_{\theta(t)}+\frac{\partial f_{\theta(t)}}{\partial \mathbf{u}}\frac{\partial \mathbf{u}}{\partial t}  \nonumber
    \\
    &\ne \lim\nolimits_{t \to t^{+}} \frac{\partial f_{\theta(t)}}{\partial \mathbf{x}}f_{\theta(t)}+\frac{\partial f_{\theta(t)}}{\partial \mathbf{u}}\frac{\partial \mathbf{u}}{\partial t}
\end{align}
}
%if we consider a switched system with two modes $f_1(\mathbf{x}(t)) = \mathbf{A}_1\mathbf{x}(t)$ and $f_2(\mathbf{x}(t)) = \mathbf{A}_2\mathbf{x}(t)$, and the event (\ref{eq:switching_x}) does not happen if the system switches at a particular state where $(\mathbf{A}_1-\mathbf{A}_2)\mathbf{x}(t) = 0$. In this case, the switching occurrence event can be defined by $\mathbf{A}_1 \ne \mathbf{A}_2$ reflected by 
\weiming{The above condition can be reflected by 
\begin{align}
    \left\| \mathbf{x}''_{+}(t) -\mathbf{x}''_{-}(t) \right\| > 0 \label{eq:switching_x_2}
\end{align}
which generates $\mathcal{S}'' = \{t \in \mathbb{R}_+ \mid   \left\| \mathbf{x}''_{+}(t) -\mathbf{x}''_{-}(t) \right\| > 0\}$, and the detected switching set is $\mathcal{S}' \bigcup \mathcal{S}'' \subseteq \mathcal{S}$. 
}

\weiming
{
The above process can be further generalized to $p$th derivative of $\mathbf{x}$ to  generate switching set $\mathcal{S}^{(p)} =  \{t \in \mathbb{R}_+ \mid \left\| \mathbf{x}^{(p)}_{+}(t) - \mathbf{x}^{(p)}_{-}(t) \right\|> 0\}$, and the detected switching set is $\mathcal{S}=\bigcup\nolimits_{i=1}^{p}\mathcal{S}^{(i)}$.
}

\weiming
{
In summary, switching occurrence events can be represented by the discontinuity of derivatives of system state $\mathbf{x}(t)$, thus the switching detection problem is converted to the problem of the discontinuity detection for derivatives of state $\mathbf{x}(t)$. The discontinuity detection of derivatives of a signal can be performed by multiple methods within the scope of signal processing, e.g., the well-known wavelet method.
}

\weiming
{
In the discrete-time case, the above discontinuity detection of derivatives of state $\mathbf{x}(t)$ can be formulated in a discretized version using $\mathbf{x}(k)$. For the $p$th derivative case,  the switching occurs at instant $k$ if the following condition is triggered
\begin{align}\label{eq:switching_d}
    \mathbf{s}^{(p)}(k)=\frac{\left\|\Delta^{(p)}\mathbf{x}(k+1)- \Delta^{(p)}\mathbf{x}(k)\right\|}{\left\|\Delta^{(p-1)}\mathbf{x}(k)\right\|} \ge \epsilon_{p}
\end{align}
where $\Delta^{(0)}\mathbf{x}(k) = \mathbf{x}(k)$ and $\Delta^{(p)}\mathbf{x}(k) =\Delta^{(p-1)}\mathbf{x}(k) - \Delta^{(p-1)}\mathbf{x}(k-1)$, and $\epsilon_{p} >0$ is  a  prescribed  threshold. For instance, (\ref{eq:switching_x})  can be rewritten as 
\begin{align}
    \mathbf{s}^{(1)}(k) =\frac{ \left\|\mathbf{x}(k+1)-2\mathbf{x}(k) + \mathbf{x}(k-1)\right\|}{\left\|\mathbf{x}(k)\right\|} \ge \epsilon_1 \label{switching_trigger}
\end{align}
where $k = 1,2\ldots K-1$.
}

\weiming
{
Based on (\ref{eq:switching_d}), the induced switching set derived from input-state traces $\phi_K^{(n)}$ can be determined by $\mathcal{S}_n=\bigcup\nolimits_{i=1}^{p}\mathcal{S}^{(i)}$ where
$
    \mathcal{S}^{(i)} = \{k \mid \mathbf{s}^{(i)}(k)  \ge \epsilon_{i},~\epsilon_{i} > 0\} \label{switching_condition}
$.
}

\begin{remark}
\weiming
{
The proposed condition (\ref{eq:switching_d}) provides a data-driven detection framework of switching occurrence events. There are two factors in the proposed framework that need some further trade-off discussions for practical applications:
\begin{enumerate}
    \item Using higher-order derivatives, i.e., larger $p$ in (\ref{eq:switching_d}), can detect more singular switching occurrence events for low-order derivatives, but the detection results are more sensitive to the noise in the measured data which may lead to faulty switching detection. Thus, a trade-off should be considered for the choice of the order of derivatives. The decision is supposed to heavily rely on the occurrence of singular switching events and the quality of collected data. 
    \item The threshold $\epsilon_{i}$ is crucial in the proposed framework. A large threshold may neglect switching occurrence events, and on the other hand, a small threshold may cause faulty switching detection. Thus, a trade-off should be considered to determine a proper threshold in the proposed framework, e.g., a threshold can be chosen small enough to include all abrupt changes but also should be larger than rest of values in $\mathbf{s}(k)$ by observation as in the DC-DC converter example in Section IV. More advanced methods of threshold selection will be our future work.
\end{enumerate}
}
\end{remark}

Considering multiple input-state traces $\phi_K^{(n)}$, $n=1,2,\ldots,N$, we can obtain $N$ switching instant sets $\mathcal{S}_n$, $n = 1,2,\ldots,N$ and the switching set for the switched system is the union of these sets, which is described as
\begin{align} \label{switching_set}
    \mathcal{S} = \bigcup\nolimits_{n=1}^{N}\mathcal{S}_n
\end{align}
where $\mathcal{S} =\{k_1,k_2,\ldots,k_\ell\}$, $0 < k_1 \le\ k_\ell < K$ includes all switching instants detected from the collected traces.

With the switching instant set $\mathcal{S}$ containing all detected switching instants, each trace $\phi_{K}^{(n)}$ can be split to $\ell$ segments $\{\phi_{K,1}^{(n)},\phi_{K,2}^{(n)},\ldots,\phi_{K,\ell}^{(n)}\}$ where $\phi_{K,m}^{(n)} = \{\mathbf{x}(k),\mathbf{u}(k)\}$, $k=k_m,\ldots,k_{m+1}$, $m = 1,2,\ldots,\ell$. Then, by consolidating the $m$th segments of $N$ traces, the following segmented training data  generated from collected $N$ traces in the $K$-length time window can be obtained
\begin{align} \label{Phi}
    \Phi_{K,m} = \{\phi_{K,m}^{(1)},\phi_{K,m}^{(2)},\ldots,\phi_{K,m}^{(N)}\},~m = 1,2,\ldots,\ell .
\end{align}

These segmented data $\Phi_{K,m}$, $m = 1,2,\ldots,\ell$ will be used for  training data-driven models of subsystems $f_i$, $i \in \mathcal{I}$.

\subsection{Subsystem Modeling}

For each segmented traces $\Phi_{K,m}$, we can extract the training data of input and output. The
input data is $N(k_{m+1}-k_m)$ pairs of $\{\mathbf{x}(k),\mathbf{u}(k)\}$ and output data is $N(k_{m+1}-k_m)$ state values $\mathbf{x}(k+1)$. Thus, the input-output training samples are
\begin{align}
    \{\tilde{\mathbf{x}}_{i,m},\mathbf{t}_{i,m}\},~i = 1,2,\ldots,\tilde N_m
\end{align}
where $\tilde{\mathbf{x}}_{i,m}=\{\mathbf{x}(k),\mathbf{u}(k)\}$,  $\mathbf{t}_{i,m} = \mathbf{x}(k+1)$, $k = k_m,\ldots,k_{m+1}$, $\tilde N_m = N(k_{m+1}-k_m)$.

Intuitively, we can use an ELM in the form of (\ref{SLFN}) to approximate system dynamics in the interval $[k_m,k_{m+1}]$ via solving the following optimization problem
\begin{align}
    \min_{\boldsymbol \upbeta \in\mathbb{R}^{L \times n_x}}\left\|\mathbf{H}_m\boldsymbol \upbeta - \mathbf{T}_m\right\|
\end{align}
where $\mathbf{H}_m$ is the randomized hidden layer output matrix based on training input $\tilde{\mathbf{x}}_{i,m}$, 
and $\mathbf{T}_m$ is the training data target matrix constructed based on $ \mathbf{t}_{i,m}$, respectively.

The above training process will generate $\ell$ optimal output  weight  vectors $\boldsymbol\upbeta_m$ for $\ell$ individual segmented traces $\Phi_{K,m}$.
This will lead to $\ell$ ELMs for $\ell$ segmented traces $\Phi_{K,m}$, $m=1,2,\ldots,\ell$, which is obviously not acceptable for the most cases of switched system modeling since the number of subsystems is often supposed to be much smaller than the switching instants. Therefore, the key issue to solve the problem of subsystem modeling is how to properly merge segmented traces for model training.  

%Taking advantage of ELM training process which is formulated as LS optimization problem in the form of (\ref{elm_opt}), the segment emerging criterion can be constructed based on the minimum LS solution of combined segmented trace data.

Considering two segmented traces $\Phi_{K,m_1}$ and $\Phi_{K,m_2}$, we merge two segmented data into one trace set $\{\Phi_{K,m_1},\Phi_{K,m_2}\}$ and the optimization problem is
\begin{align}
    \min_{\boldsymbol \upbeta \in\mathbb{R}^{L \times n_x}}\left\|\mathbf{H}_{m_1,m_2}\boldsymbol \upbeta - \mathbf{T}_{m_1,m_2}\right\|
\end{align}
where $\mathbf{H}_{m_1,m_2}$ is the randomized hidden layer output matrix based on the combined training input $\tilde{\mathbf{x}}_{i,m}$ as
\begin{align} \label{H12}
   \mathbf{H}_{m_1,m_2} = 
   \begin{bmatrix} 
   \mathbf{H}_{m_1}
   \\
   \mathbf{H}_{m_2}
   \end{bmatrix} 
\end{align} 
and $\mathbf{T}_{m_1,m_2}$ is the  combined training data target matrix
\begin{align}  \label{T12}
    \mathbf{T}_{m_1,m_2}  = 
    \begin{bmatrix}
    \mathbf{T}_{m_1}
    \\
    \mathbf{T}_{m_2}
    \end{bmatrix} .
\end{align}

Solving the above LS optimization problem to obtain the solution is $\boldsymbol\upbeta_{m_1,m_2}^{*}$, and the minimal value is
\begin{align}
    \gamma (\boldsymbol\upbeta_{m_1,m_2}^{*}) =  \left\|\mathbf{H}_{m_1,m_2}\boldsymbol \upbeta_{m_1,m_2}^{*} - \mathbf{T}_{m_1,m_2}\right\| .
\end{align}

If $\gamma (\boldsymbol\upbeta^{*}_{m_1,m_2})$ is sufficient small, i.e., smaller than a prescribed threshold $\zeta > 0$, it implies that there exist one single common output weight vector $\boldsymbol\upbeta^{*}_{m_1,m_2}$ that works for both segmented data $\Phi_{K,m_1}$ and $\Phi_{K,m_2}$ with a small approximation error less than the prescribed threshold $\zeta$, leading to a common ELM model extracted from two segmented data. 

The segmented data merging process can be performed recursively  from $\Phi_{K,1}$ to $\Phi_{K,\ell}$ to eventually properly merge all segmented data and model subsystem dynamics. The process is summarized in Algorithm \ref{alg3}.

\begin{remark}
\weiming
{
In Algorithm \ref{alg3}, an inappropriately large threshold would wrongly merge segmented data generated from different subsystems into one trace set, and on the other hand,  an inappropriately small threshold may fail to merge segmented data from one same subsystem into one trace set. Thus, the prescribed threshold $\zeta$ in the segmented data merging and subsystem modeling process is crucial for the performance of Algorithm \ref{alg3}. Developing advanced methods of threshold selection will be future work to enrich the modeling framework proposed in this paper.
}
\end{remark}

\begin{algorithm}[ht!]
\SetAlgoLined
\SetKwInOut{Input}{Input}
\SetKwInOut{Output}{Output}
\SetKw{Return}{return}
\Input{Segmented trace data $\Phi_{K,m}$, $n=1,\ldots,\ell$}
\Output{Subsystem ELM models $f_i$, $i \in \tilde{\mathcal{I}}$}
\Fn{$\mathrm{systemModel}(\Phi_{K,m})$}{

$n \gets \ell$, $m_1 \gets 1$;

\tcc{Segmented data merge}

\While{$m_1 < n$}
{
$m_2 \gets 1$;

\While{$m_2 \le n$}
{$m_2 \gets m_2 + 1$; 

%Solve 
$\min_{\boldsymbol \upbeta \in\mathbb{R}^{L \times n_x}}\left\|\mathbf{H}_{m_1,m_2}\boldsymbol \upbeta - \mathbf{T}_{m_1,m_2}\right\|$;

\If{$\gamma(\boldsymbol\upbeta^{*}_{m_1,m_2}) \le \zeta$}
{
$\Phi_{K,m_1} \gets \{\Phi_{K,m_1},\Phi_{K,m_2}\}$; 

$n \gets n - 1$;
}

}
$m_1 \gets m_1 + 1$;
}
\tcc{Generate subsystem models}
$m \gets 1$;

\While{$m \le m_1$}
{%Solve 
$\min_{\boldsymbol \upbeta \in\mathbb{R}^{L \times n_x}}\left\|\mathbf{H}_{m}\boldsymbol \upbeta - \mathbf{T}_{m}\right\|$;

Obtain solution $\boldsymbol\upbeta^{*}_{m}$, $m=1,\ldots,m_1$ to construct subsystem models $f_i$, $i=1,\ldots,m_1$ in the form of (\ref{SLFN});

\Return{subsystem models $f_i$, $i=1,\ldots,m_1$}
}
}
 \caption{Subsystem Modeling} \label{alg3}
\end{algorithm}

\subsection{Switching  Reconstruction}

As the input-state traces $\phi_K^{(n)}$, $n = 1,2\ldots,N$ collected in a finite-time window with a length of $K$ steps, we can infer time-dependent switching laws valid for the $K$-length time interval. Thus, the time-dependent switching law reconstruction for switching signal $\theta(k)$, $k \in \mathbb{N}$ is reduced to reconstruction of time-dependent function $\theta(k)$, $k \in [0,K]$ where $K$ is the length of training traces. 

Using the switching detection results and subsystem modeling results by Algorithm \ref{alg3}, we can label the time segments obtained by switching detection with corresponding subsystem index obtained by Algorithm \ref{alg3} to generate the following sequence 
\begin{align}
    \{\theta(k_m-1),\theta(k_m)),k_m\},~m = 1,2,\ldots,\ell
\end{align}
where $k_m$, $m = 1,2,\ldots,\ell$ is the detected switching instants, $\theta(k_m-1)$ and $\theta(k_m)$ indicate the system switching from index $\theta(k_m-1)$ to $\theta(k_m)$. 

\weiming{If more information is provided, e.g., the switching law is a periodic or semi-periodic switching, the switching law $\theta(k)$ can be inferred based on the traces in the $K$-length time interval.} A semi-periodic switching law reconstruction is illustrated by the data-driven modeling of DC-DC converters in the next section.

\section{Evaluation by Data-Driven Modeling of DC-DC Converters}

\begin{figure}
\centering
	\includegraphics[width=4cm]{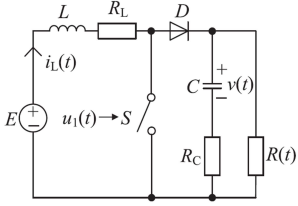}
	\caption{DC-DC boost converter.}
	\label{fig:dcdc} 
\end{figure}

\weiming{The DC-DC converter is a typical switched system. %Power converters are widely-used industrial devices and the switching lies in the abrupt change between different electrical circuit structures to transform a constant or slowly varying DC voltage to a DC voltage independent of the load.
The example presented is borrowed from \cite{heemels2009introduction}, which is a boost converter with its topology shown in Fig. \ref{fig:dcdc}. The values of these circuit components are listed in Table \ref{tab1}. %The system has to be stabilized at a limit cycle rather than in an equilibrium state.
When the switch is \emph{on}, there is an increase in the inductor current $i_L(t)$. On the other hand, when the switch is \emph{off}, the current ramps down through the
flyback diode, the capacitor, and the load. This leads to a transfer of the energy accumulated in the inductance into the capacitor. The process executes cyclically, whereas the boost converter operates with the switching period $T_s$ and the duty cycle $u \in [0,1]$, which corresponds
to the ratio of activation duration of an \emph{switch-on} mode to the period.}

\begin{table}[h!] 
\centering
\caption{Values of circuit components}
\label{tab1}
\begin{tabular}{ c|ccccccc }
\hline
Component & $E$ & $L$ & $R_L$ & $C$ & $R_C$ & $T_s$
\\
\hline
Value & $20$ V & $1$ mH & $0.1$ $\Omega$ & $10$ $\mu$F  & $0.06$ $\Omega$ & $0.1$ ms 
\\
\hline
\end{tabular}
\end{table}

Corresponding to the modes of \emph{switch-on} and \emph{switch-off}, the DC-DC model can be constructed. Letting system state $\mathbf{x}(t) = [i_{L}(t)~~ v(t)]^{\top}$, the state-space model of the DC-DC converter is in the following form 
\begin{align} \label{dcdc}
    \dot{\mathbf{x}}(t) = \mathbf{A}_{\theta(t)}\mathbf{x}(t) + \mathbf{B}_{\theta(t)}
\end{align}
where $\theta(t) = {1,2}$ is the switching law defined by
\begin{align} \label{theta}
    \theta(t) = \begin{cases}
    1,~t \in [2nT_s,(2n+1)T_s+uT_s)
    \\
    2,~t \in [(2n+1)T_s+uT_s,2(n+1)T_s)
\end{cases}
\end{align}
in which $n= 0,1,2,\ldots$. System matrices are given as below:
\begin{align*}
    &\mathbf{A}_1 = 
    \begin{bmatrix}
    {-\frac{R_L}{L}} & {0}
    \\
    {0} & {\frac{1}{(R+R_C)C}}
    \end{bmatrix},~
    \mathbf{B}_1 = 
    \begin{bmatrix}
    \frac{E}{L}
    \\
    0
    \end{bmatrix}
    \\
     &   \mathbf{A}_2 = 
    \begin{bmatrix}
    {-\frac{R_L}{L}-\frac{R_CR}{L(R+R_C)}} & {-\frac{R}{R+R_C}L}
    \\
    {\frac{R}{R+R_C}C} & {-\frac{1}{R+R_C}C}
    \end{bmatrix},~
    \mathbf{B}_2 = 
    \begin{bmatrix}
    \frac{E}{L}
    \\
    0
    \end{bmatrix} .
\end{align*}

To evaluate the proposed data-driven modeling method for switched systems, we first collect system traces generated by the DC-DC model, then utilize the modeling framework proposed in this paper to develop a data-driven ELM model 
\begin{equation}
\mathbf{x}(k+1)=f_{\theta(k)}\left(\mathbf{x}(k)\right)
\end{equation}
where $f_i$ are ELMs and expected to be able to effectively predict the system behaviors of the DC-DC converter. 

\begin{figure}
\centering
	\includegraphics[width=7cm]{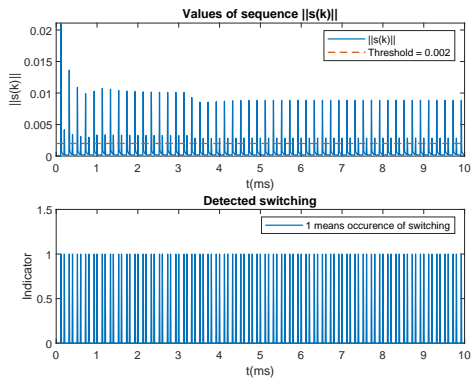}
	\caption{Switching detection.}
	\label{fig:switching} 
\end{figure}

\begin{figure}
\centering
	\includegraphics[width=7cm]{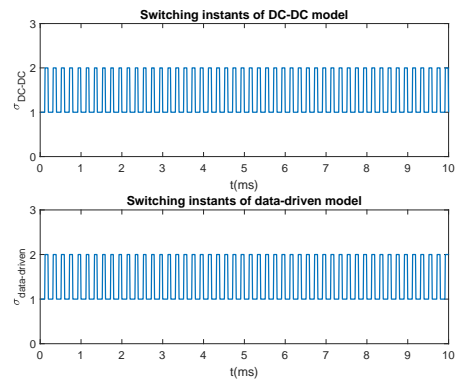}
	\caption{Reconstructed switching instants.}
	\label{fig:instant} 
\end{figure}

With a sampling period of $0.01$ ms, we collect  20000 training data from 20 system traces in a $10$ ms time window generated from the DC-DC model (\ref{dcdc}), then the data-driven modeling process is illustrated below\footnote{The source code is available at: \url{https://github.com/xiangweiming/data-driven-modeling-switched-systems}}:

\begin{enumerate}
    \item \textbf{Switching Detection:} By executing switching detection with a  prescribed \weiming{threshold $\epsilon_1 = 0.002$} for the average value of $\mathbf{s}(k)$ of 20 traces, the switching instants are detected as shown in Fig. \ref{fig:switching}. Training data is also segmented for subsystem modeling.
    
    \item \textbf{Subsystem Modeling:} Using Algorithm \ref{alg3} with a prescribed threshold $\zeta = 1$ and ELM models with 200 neurons and {\tt sigmoid} activation functions, two subsystems of ELM are obtained.  
    \item \textbf{Switching Law Reconstruction:}
    Based on the obtained switching instants and two subsystems, the switching sequence     $\{\theta(k_m-1),\theta(k_m)),k_m\},~m = 1,2,\ldots,\ell$ is reconstructed for interval $[0,10]$ ms as shown in Fig. \ref{fig:instant}, which can successfully reconstruct the switching signal $\theta(k)$ described by (\ref{theta}). Given the prior information that the switching signal is a semi-periodic switching between subsystems, the switching law can be reconstructed as 
    \begin{align*} 
    \theta(k) = \begin{cases}
    1,~k \in [20n,20n+11]
    \\
    2,~k \in [20n+11,20n+19] 
\end{cases}
\end{align*}
where $n=0,1,\ldots$. 
\end{enumerate}

\begin{figure}
\centering
	\includegraphics[width=7cm]{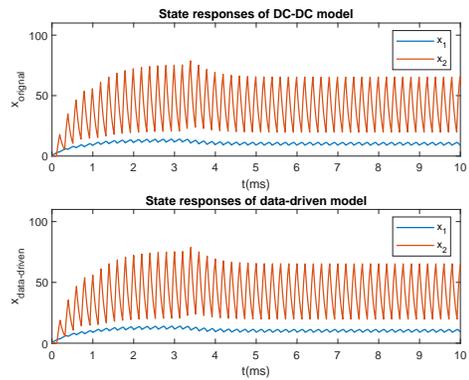}
	\caption{State responses of DC-DC model and data-driven model.}
	\label{fig:response} 
\end{figure}

\begin{figure}
\centering
	\includegraphics[width=7cm]{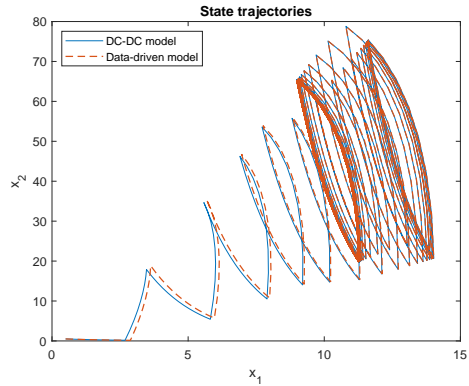}
	\caption{State trajectories of DC-DC model and data-driven model.}
	\label{fig:trajectory} 
\end{figure}

Given the initial state $\mathbf{x}_0 = [0.5~~0.5]^{\top}$, the state responses and trajectories of data-driven switched system model are shown in Figs. \ref{fig:response} and \ref{fig:trajectory} which show data-driven model can predict the system trajectories very well. 

\section{Conclusion and Further Remarks}

\weiming{A data-driven modeling framework of switched dynamical systems using ELM has been studied in this paper.} An switching detection method is developed to identify the time instants when switching occurs and generate segmented training data. ELM is then utilized to learn subsystem dynamics. The modeling algorithm is able to properly merge segmented data and  learn  subsystem models. Finally, the switching sequence with both detected switching instants and subsystem activation indices is reconstructed. The developed approach is evaluated by an example of DC-DC converters.   

\weiming{
Within the modeling framework of using ELM, there exist several challenging problems remain open for further study. For instance, as aforementioned, the selection of thresholds in switching detection and data merging is crucial in the proposed modeling framework and deserves an in-depth study. In addition, this paper assumed that the data has been well-denoised before modeling, it is worthwhile to further explore the potentials of the proposed modeling framework in presence of innegligible noise in measured data. It is also meaningful to extend the proposed framework to state-dependent switched systems in future work.
}

\bibliographystyle{ieeetr}
\bibliography{ref}

\end{document}